\documentclass[reprint,amsmath,amssymb,aps,floatfix]{revtex4-2}
\usepackage{graphicx}
\usepackage{dcolumn}
\usepackage{bm}
\usepackage{hyperref}
\usepackage{lipsum}
\usepackage{subcaption}  
\usepackage{caption}         
\usepackage{float}
\usepackage{tikz}
\usepackage{ragged2e}
\usepackage{placeins}
\usetikzlibrary{positioning, arrows.meta}  
\usepackage{relsize}  
\usepackage{overpic}  
\usepackage{placeins}
\usepackage{amssymb}
\usepackage{amsmath}
\usepackage{algorithm}
\usepackage{algpseudocode}
\usepackage{makecell}
\usepackage{bm}
\usepackage{placeins}
\usepackage[export]{adjustbox}
\usepackage{amsthm}

\theoremstyle{remark}

\usepackage{mathtools}
\usepackage{float}
\usepackage{tikz}
\usepackage{ragged2e}
\usetikzlibrary{positioning, arrows.meta}  
\usepackage{relsize}  
\usepackage{overpic}

\usepackage{tabularx}
\usepackage{booktabs}
\usepackage{array}
\usepackage{enumitem}
\usepackage{ragged2e}
\usepackage{makecell}
\begin{document}

\preprint{APS/123-QED}

\title{At the Top of the Mountain, the World can Look Boltzmann-Like:\\ Sampling Dynamics of Noisy Double-Well Systems}

\author{Abir Hasan, Nikhil Shukla}
\affiliation{%
University of Virginia, Charlottesville, VA, USA,
}%

\begin{abstract}
The success of the transistor as the cornerstone of digital computation motivates analogous efforts to identify an equivalent hardware primitive—the probabilistic bit (p-bit)—for the emerging paradigm of probabilistic computing. Here, we uncover a fundamental \emph{ubiquity} in the stochastic dynamics of double-well energy systems when initialized near the barrier top. Using a topological framework grounded in Morse and singularity theory, we make use of the result that all smooth, even double-well potentials reduce near the saddle point to a canonical quartic normal form. Within this regime, the interplay of noise, synaptic bias, and potential curvature produces a topologically robust short-time evolution characterized by a $\tanh$-like response—enabling Boltzmann-like sampling that is largely independent of the detailed shape of the potential, apart from its effective temperature scaling. Analytical derivations and numerical simulations across multiple representative systems corroborate this behavior. Our work provides a unifying foundation for assessing and engineering a broad class of physical platforms—including oscillators, bistable latches, and magnetic devices—as p-bits operating within a \emph{synchronous} framework for stochastic sampling and probabilistic computation.

\end{abstract}
                             
\maketitle

\section{Introduction}

Stochastic sampling underpins a wide range of applications, from training energy-based machine-learning models such as restricted Boltzmann machines (RBMs) to probabilistic inference in scientific modeling and risk analysis in finance~\cite{lye2019review,pavliotis2014stochastic}. While the algorithmic foundations of stochastic sampling have been extensively developed, its hardware realization has only recently gained attention. This is because stochastic sampling, which fundamentally relies on random fluctuations, lacks a natural physical substrate analogous to CMOS for deterministic digital logic. This gap began to close with the advent of the \emph{probabilistic bit} (p-bit) concept, first realized using low-barrier magnetic tunnel junctions (MTJs)~\cite{camsari2017implementing}. Magnetic tunnel junctions (MTJs) constitute a canonical double-well energy system that, when engineered with sufficiently low barriers, exhibits thermally activated state transitions whose probabilities follow the Boltzmann distribution~\cite{faria2017low}.

P-bits act as tunable random number generators whose output statistics can be continuously biased by an input signal, providing a natural hardware building block for stochastic samplers. This property has enabled p-bits to serve as physical primitives for a range of applications—from solving combinatorial optimization problems (COPs) to AI training and implementing invertible logic~\cite{chowdhury2023full,camsari2017stochastic}. In parallel, a diverse range of hardware technology platforms—including ferroelectrics~\cite{luo2023probabilistic}, phase-transition oxides (e.g., $\text{VO}_{\text{2}}$~\cite{deng2023hydrogenated} , $\text{NbO}_{\text{2}}$~\cite{rhee2023probabilistic}), and CMOS-based relaxation oscillators~\cite{aadit2022physics}—have been explored as potential p-bit implementations.

Traditionally, p-bits operate in an \emph{asynchronous mode}, where the system stochastically switches between two stable states, and the relative dwell times in each state determine the output probability. These dwell times are governed by the \emph{effective} energy barrier height, which can be modulated by the synaptic input—exhibiting temporal behavior generally consistent with Kramers’ escape dynamics~\cite{adeyeye2025sampling,brown1963thermal,daniels2023neural}. 

More recently, Ekanayake \emph{et al.}~\cite{ekanayake2025bridging} showcased an alternative mode of operation based on the dynamics of an oscillator under second-harmonic injection (SHI)~\cite{ekanayake2025bridging}. In this regime, the system is periodically driven to the \emph{top of the energy barrier}, and the ensuing short-time stochastic relaxation can encode the Boltzmann distribution. The dynamics are then governed by the gradient-flow properties in the presence of noise, modulated by the local curvature of the potential near the saddle point. Similarly, other works~\cite{ostwal2018spin,debashis2020hardware} have leveraged this operating mode in spintronic devices. Unlike traditional asynchronous operation, this approach can enable synchronous operation wherein a global timing signal dictates when the system is driven to the barrier top, and the sampling event completes after a short incremental time $\Delta t$ once the trajectory irreversibly falls into one of the wells. 

Motivated by these observations, we examine the extent to which this behavior may be ubiquitous; that is, to what degree any double-well system initialized at the barrier top is capable of approximating Boltzmann-like sampling. Many hardware systems—including magnetic tunnel junctions, analog oscillators under second-harmonic injection (SHI), and CMOS bistable latches—naturally exhibit such double-well energy landscapes. Demonstrating that this behavior is ubiquitous could provide a common computational fabric for employing these and other physical systems that realize double-well energy landscapes as p-bits in synchronous probabilistic computing architectures

\section{Realizing Boltzmann Sampling Through System Dynamics}
As established by Camsari \emph{et al.}~\cite{camsari2017stochastic}, a fundamental requirement for designing a p-bit capable of sampling from the Boltzmann distribution is that the output state is updated as,
\begin{equation}
    \sigma^{+} = \operatorname{sgn}\!\big[\tanh(\beta h) - \eta\big],
    \label{eq:gibbs_update}
\end{equation}
with $\sigma^+ \in \{-1,+1\}$ representing the output state, $h$ denoting the input field or synaptic bias, $\beta$ an effective inverse temperature, and $\eta$ is a random variable uniformly distributed in $[-1,+1]$ that represents stochastic fluctuations. In this expression, $\tanh(\beta h)$ captures the deterministic tendency of the system to align with its local field ($h$), while the noise term introduces probabilistic switching. 

As alluded to earlier, one method to realize such stochastic sampling behavior in a physical system is to engineer the system dynamics to be,
\begin{equation}
    s(t) \approxeq\, \tanh(\beta h t)
    \quad\Longrightarrow\quad
    \dot{s} = \beta h(1-s^{2}),
    \label{eq:logistic_drift}
\end{equation}
where, $s$ is an external observable that represents the system's measurable output. The ensuing short-time ($\Delta t$) relaxation in the presence of noise, which constitutes the sampling event, then determines which potential well the system ultimately settles into. 

We now evaluate how the noisy gradient descent dynamics ---directly tied to the energy landscape--- can facilitate such behavior. While \textit{s} is the external observable that facilitates stochastic sampling, it may not directly correspond to the internal state variable ($\vartheta)$ that parametrizes the system's potential landscape. For example, in an oscillator operating under second-harmonic injection (SHI), the internal state corresponds to the oscillator phase \( \varepsilon \), referenced such that \( \varepsilon = 0 \) aligns with the top of the potential barrier. The measurable output voltage is given by
\(
v_0 = \cos\!\left(\frac{\pi}{2} + \varepsilon\right) = -\sin(\varepsilon),
\)
which yields the mapping \( s = f(\vartheta) = -\sin(\varepsilon) \) between the internal coordinate and the external observable output. The details of oscillator system have been described in our prior work~\cite{ekanayake2025bridging}.\\

\noindent \textbf{Formulating the $s=f(v)$ map:}  
For gradient-flow dynamics described by \(\dot{\vartheta}=g(\vartheta)=-U'(\vartheta)\), 
where \(U(\vartheta)\) denotes the potential energy landscape, 
the corresponding mapping to the observable \(s\), designed to evolve as \(s=\tanh(\beta h t)\) 
to realize Boltzmann sampling, can be expressed as,
\begin{equation}
    f(\vartheta)
    \;=\;
    \tanh\!\left(
        \beta h \int \frac{d\vartheta}{g(\vartheta)}
    \right)
    \label{eq:map_eq}
\end{equation}

\noindent
thereby relating the system coordinate \(\vartheta\) to its effective stochastic state variable \(s\).
A detailed derivation of Eq.~\eqref{eq:map_eq} is provided in Appendix~\ref{ap:s_map}.\\

\noindent \textbf{Implications of the choice of map for stochastic sampling:}  
To evaluate the impact of the mapping $f(\vartheta)$, we consider the gradient-flow dynamics in the presence of stochastic fluctuations which can be expressed as,
\begin{equation}
    \,d \vartheta = g(\vartheta) \,dt + \sqrt{2K_n}\,dW_t,
    \label{eq:langevin}
\end{equation}
where \(K_n\) denotes the noise intensity.  
Expressing Eq.~\eqref{eq:langevin} in terms of the observable variable \(s=f(\vartheta)\) yields,
\begin{equation}
    \,ds = f'(\vartheta)g(\vartheta) \,dt + f'(\vartheta)\sqrt{2K_n}\,dW_t.
    \label{eq:transformed_dynamics}
\end{equation}
In practical systems, if the mapping \(f(\vartheta)\) is chosen such that the effective drift term \(f'(\vartheta)g(\vartheta)\) near the barrier top~\emph{approximates} the symmetric logistic form of Eq.~(\ref{eq:logistic_drift}), the observable dynamics in \(s\) can emulate Boltzmann sampling. 

The specific form of \(f(\vartheta)\), however, determines how the internal noise is projected into the observable domain. Because the effective noise amplitude scales as \(f'(\vartheta)\), the mapping directly influences the \emph{effective temperature} of the sampling process:
\[
T_{\mathrm{eff}} \propto K_n\,\langle [f'(\vartheta)]^{2}\rangle
\]
for a given intrinsic noise strength \(K_n\). This insight highlights the mapping \(f(\vartheta)\) as a potential \emph{design degree of freedom}: by tailoring the transduction between the internal state variable and its observable representation, one can effectively tune the sampling temperature without modifying the underlying noise source.

\section{Illustrative Examples}
\subsection{Harmonic Oscillator under First \& Second Harmonic Injection}

As an illustrative example of a double-well energy system that admits a well-defined map, \(s=f(\vartheta)\), 
we consider a harmonic oscillator subject to simultaneous first- and second-harmonic injection. 
In the rotating frame of reference, the effective potential of such an oscillator can be expressed as
\begin{equation}
    U_{\mathrm{osc}}(\varepsilon)
    \;=\;
    \gamma \sin(\varepsilon)
    \;+\;
    \tfrac{1}{2}K_s \cos(2\varepsilon),
\end{equation}
where \(\varepsilon\) denotes the oscillator phase, referenced such that 
the potential minima occur at \(\varepsilon = \pm \tfrac{\pi}{2}\) 
and the energy maximum at \(\varepsilon = 0\). 
Here, the first term (\(\gamma \sin\varepsilon\)) represents the effect of 
\emph{first-harmonic injection}, which biases the oscillator phase in proportion 
to the injected signal amplitude \(\gamma\,\,(\equiv\,h)\). 
The second term (\(K_s \cos(2\varepsilon\))) corresponds to \emph{second-harmonic injection} (SHI), 
which establishes a symmetric double-well potential and facilitates bistable phase locking. 
The parameter \(\gamma \in \mathbb{R}\) acts as the effective synaptic input: 
\(\gamma>0\) corresponds to in-phase injection, while \(\gamma<0\) represents out-of-phase injection. 
The oscillator phase \(\varepsilon\) thus serves as the internal state variable 
\((\varepsilon \equiv \vartheta)\).

The corresponding gradient-flow dynamics of the oscillator are given by \cite{ekanayake2025bridging}
\begin{equation}
    \dot{\varepsilon}
    \;=\;
    -\gamma\cos(\varepsilon)
    \;+\;
    K_s\sin(2\varepsilon),
    \label{eq:osc}
\end{equation}

\noindent To define a valid observable mapping, we set
\begin{equation}
s_{\mathrm{osc}} = \cos\!\left(\tfrac{\pi}{2}+\varepsilon\right)
= -\sin(\varepsilon), \label{eq:osc_map1}   
\end{equation}
Substituting Eq.~\eqref{eq:osc_map1} into Eq.~\eqref{eq:osc} yields the transformed dynamics
\begin{equation}
    \dot{s}_{\mathrm{osc}}
    = (1-s_{\mathrm{osc}}^{2})(\gamma + 2K_s s_{\mathrm{osc}}),
    \label{eq:osc_map2}
\end{equation}
where the prefactor \((1-s_{\mathrm{osc}}^{2})\) ensures bounded evolution in the observable domain.

In the weak second-harmonic limit (\(K_s \ll 1\)), which corresponds to a small energy barrier, Eq.~\eqref{eq:osc_map2} simplifies to,
\begin{equation}
\dot{s}_{\mathrm{osc}} \;\approx\; \gamma(1-s_{\mathrm{osc}}^{2})
\quad \Longrightarrow \quad
s_{\mathrm{osc}}(\Delta t)
= \tanh(\gamma\Delta t),    
\end{equation}
indicating that, over short relaxation intervals \(\Delta t\) (see Appendix~\ref{ap:delta_t} ), the oscillator’s drift dynamics exhibit the canonical \(\tanh(\cdot)\) dependence required for Boltzmann sampling (Eq.~\eqref{eq:gibbs_update}). 

\begin{table*}[t] 
  \centering
  \caption{Canonical formulations of smooth double-well potentials.}
  \label{tab:potentials}
  \setlength{\tabcolsep}{22pt}   
  \renewcommand{\arraystretch}{1.5} 
  \begin{tabularx}{\linewidth}{c *{2}{c}} 
    \toprule
    Formulation & Potential function $U(\vartheta)$ & Condition for Double Well  \\
    \midrule
    \makecell{Even Polynomials} & 
    \makecell{$U(\vartheta) = \sum_{k=2}^{n} a_{2k} \vartheta^{2k} - b \vartheta^2$ \\\\ Parameters used in simulations \\ $a_2=-0.5;a_4=0.5;a_6=0.05$} & 
    \makecell{$a>0; b>0$\\ sufficient condition\\ (see Appendix~\ref{ap:DW})} \\ [35pt]

    \makecell{Double-Gaussian Wells\\(centered at $\pm \vartheta_{0}$)} & \makecell{$U(\vartheta) = -A \!\Bigg[
    e^{-\dfrac{(\vartheta\,-\,\vartheta_{0})^{2}}{2\sigma^{2}}}
    + e^{-\dfrac{(\vartheta\,+\,\vartheta_{0})^{2}}{2\sigma^{2}}}
    \Bigg]$ \\\\Parameters used in simulations\\$A=1;  \sigma=0.25; \vartheta_{0}=0.5$} & \makecell{$A>0$\,;\,$\dfrac{\vartheta_{0}}{\sigma} > 1,$}\\ [45pt]

    \makecell{Exponential tails} & \makecell{$U(\vartheta)=A\big(\cosh(\beta \vartheta)-1\big)-b\,\vartheta^{2}$ \\\\ Parameters used in simulations \\ $A=1; \beta=1; b=2.6$} & \makecell{$b>\tfrac{1}{2}A\beta^{2}$ \\negative quadratic near $0$,\\ exponential growth at $\lvert \vartheta\rvert\to\infty$.} \\

    \bottomrule
  \end{tabularx}
\end{table*}

\subsection{Magnetic Tunnel Junction (MTJ)}

As a second example, we consider an archetypal magnetic tunnel junction (MTJ), whose energy landscape also exhibits 
a double-well profile. The total magnetic energy of the free layer can be expressed as,~\cite{guimaraes2009principles}
\begin{equation}
    U_{\mathrm{MTJ}}(\theta)
    \;=\;
    K_u V \sin^2(\theta)
    \;-\;
    \mu_0 M_s V H \cos(\theta),
    \label{eq:MTJ_E}
\end{equation}
where the parameters are defined as follows: 
\(U_{\mathrm{MTJ}}\) is the total magnetic energy of the free layer; 
\(K_u\) is the uniaxial anisotropy constant; 
\(V\) is the volume of the free magnetic layer; 
\(\mu_0\) is the permeability of free space; 
\(M_s\) is the saturation magnetization; 
\(H\) is the \emph{external} magnetic field applied along the easy (\(z\)) axis; 
and \(\theta\) denotes the angle between the magnetization vector and the easy axis. In Eq.~\eqref{eq:MTJ_E}, the two stable minima are located at \(\theta = \{0,\pi\}\), 
corresponding to parallel and antiparallel magnetic orientations, while the energy barrier maximum occurs at 
\(\theta = \tfrac{\pi}{2}\). 

To align the coordinate system with the top of the barrier, we introduce an offset
\(
\vartheta_{\mathrm{MTJ}} = \tfrac{\pi}{2} + \theta,
\)
which shifts the energy maxima to \(\vartheta_{\mathrm{MTJ}} = 0\) 
and the minima to \(\vartheta_{\mathrm{MTJ}} = \pm \tfrac{\pi}{2}\). 
In this shifted frame, the internal state variable is represented by \(\vartheta_{\mathrm{MTJ}}\), 
while the magnetization component along the easy axis 
(\(\cos\theta \equiv -\sin\vartheta_{\mathrm{MTJ}}\)) serves as the observable output.

The potential function in this transformed frame then becomes,
\begin{equation}
\begin{split}
    U_{\mathrm{MTJ}}(\vartheta_{\mathrm{MTJ}})
    &= K_u V \cos^2(\vartheta_{\mathrm{MTJ}})
    + \mu_0 M_s V H \sin(\vartheta_{\mathrm{MTJ}}) \\\\
    &\equiv a \cos^2(\vartheta_{\mathrm{MTJ}}) + b \sin(\vartheta_{\mathrm{MTJ}}),
    \label{eq:MTJ_En_tr}
\end{split}
\end{equation}
where \(a = K_u V\) and \(b = \mu_0 M_s V H\) for compactness. The corresponding gradient-flow dynamics can then be expressed as,
\begin{equation}
    \dot{\vartheta}_{\mathrm{MTJ}}
    = -U'_{\mathrm{MTJ}}(\vartheta_{\mathrm{MTJ}})
    = -b\cos(\vartheta_{\mathrm{MTJ}}) + a\sin(2\vartheta_{\mathrm{MTJ}}),
    \label{eq:MTJ_grad}
\end{equation}
which is analogous in form to Eq.~\eqref{eq:osc} for the oscillator under SHI with $a\equiv K_s$ and $b \equiv \gamma$.

To define the observable, we adopt the same mapping used earlier for the oscillator,
\(s_{\mathrm{MTJ}} = -\sin(\vartheta_{\mathrm{MTJ}}),\)
which corresponds directly to the magnetization projection along the easy axis.  
Applying this transformation to Eq.~\eqref{eq:MTJ_grad} yields,
\begin{equation}
    \dot{s}_{\mathrm{MTJ}}
    = (1-s_{\mathrm{MTJ}}^{2})(b + 2a s_{\mathrm{MTJ}}),
    \label{eq:MTJ_grad_tanh}
\end{equation}
which is mathematically equivalent to Eq.~\eqref{eq:osc_map2} for the oscillator. In low-barrier magnets, the anisotropy term is intentionally designed to be small (\(|a|\ll1\)); 
thus, Eq.~\eqref{eq:MTJ_grad_tanh} reduces to
\begin{equation}
\begin{split}
&\dot{s}_{\mathrm{MTJ}} \;\approx\; b(1-s_{\mathrm{MTJ}}^{2})\\\\
\Rightarrow \quad
&s_{\mathrm{MTJ}}(\Delta t)
= \tanh(b\,\Delta t)
= \tanh((\mu_0 M_s V H)\,\Delta t)
\end{split}
\end{equation}
Similar to the oscillator case, during short relaxation intervals \(\Delta t\), 
the drift component of the MTJ dynamics—initiated near the top of the barrier—exhibits the canonical 
\(\tanh(\cdot)\) dependence to facilitate Boltzmann sampling.

\section{Generalization to Smooth Even Double-Well Potentials}

Having established two physical realizations of double-well systems—the oscillator under harmonic injection, and the MTJ—we now examine how Boltzmann-like sampling dynamics can emerge in a broader class of smooth, even double-well potentials. Such potentials are characterized by two (and only two) symmetrically located energy minima at \( \pm x^{\star} \) and a non-degenerate maximum at \(x=0\). Although infinitely many analytic forms can generate such landscapes, representative examples are summarized in Table~\ref{tab:potentials}.

\subsection{Canonical Double Well Energy System}

We begin by analyzing the class of smooth, even double-well potentials through the lens of \emph{Morse theory} and \emph{singularity theory}~\cite{milnor1963morse,arnold1992singularity,golubitsky1985singularities,poston1978catastrophe}.  
According to \emph{Morse theory}, any smooth function \( U(x) \) with a nondegenerate critical point (\( U'(0)=0,\; U''(0)\neq 0 \)) can be locally transformed, via a smooth coordinate change \(x\mapsto\vartheta(x)\), into a purely quadratic form,
\[
U(\vartheta) = U(0) - \tfrac{1}{2}\lambda\,\vartheta^2,
\]
where \( \lambda = -\,U''(0) \). A positive \(\lambda\) corresponds to a local maximum (relevant to the proposed approach), while a negative \(\lambda\) corresponds to a local minimum.

We now consider a smooth, even potential (\(U(x)=U(-x)\)) with a nondegenerate maximum at the origin (\(U'(0)=0,\; U''(0)<0\)). The Taylor expansion about \(x=0\) takes the form
\begin{equation}
\begin{split}
    U(x) &= U(0) + \frac{U''(0)}{2!}x^2 + \frac{U^{(4)}(0)}{4!}x^4 + \mathcal{O}(x^6) \\\\
          &\equiv U(0) - \frac{r\,r_0}{2}x^2 + \frac{b_0}{4}x^4 + \mathcal{O}(x^6),
\end{split}
\end{equation}
where \( r\,r_0 > 0 \) ensures a local maximum at the origin. Neglecting higher-order \(\mathcal{O}(x^6)\) terms, the condition \(b_0>0\) guarantees that \(U(x)\) is bounded from below and forms a symmetric double well. Introducing rescaled variables 
\(
\vartheta = \sqrt{\tfrac{b_0}{r_0}}\,x
\)
and 
\(
\alpha = \tfrac{r_0^2}{b_0},
\)
the potential assumes the canonical quartic form characteristic of the $A_3$ (pitchfork) singularity in catastrophe theory:
\begin{equation}
U(\vartheta) = \alpha\!\left(\frac{\vartheta^4}{4} - \frac{r\, \vartheta^2}{2}\right).
\label{eq:s_pitch_fork}
\end{equation}

\noindent
Here \(r>0\) quantifies the curvature at the barrier: at \(r=0\), the single critical point becomes degenerate, and for \(r>0\) the potential bifurcates into two symmetric minima, corresponding to a stable double-well configuration. Furthermore, as shown in Appendix~\ref{ap:DW}, the system retains exactly two minima even there are higher-order even powers \(\vartheta^{2k}\) (\(k\ge3\)) with non-negative coefficients and at least one is strictly positive.

Including a weak external field \(h\in\mathbb{R}\) introduces an asymmetry by tilting the landscape:
\begin{equation}
U(\vartheta)
= \frac{\alpha}{4}\vartheta^4 - \frac{\alpha r}{2}\vartheta^2 - h\,\vartheta.
\label{eq:A3_Energy}
\end{equation}
The external bias \(h\) breaks the intrinsic \(\mathbb{Z}_2\) symmetry (\(U(\vartheta)\neq U(-\vartheta)\)), lowering one well relative to the other. The associated gradient-flow dynamics can then be expressed as,
\begin{equation}
\dot{\vartheta}=-U'(\vartheta)=h+\alpha r\vartheta-\alpha\vartheta^3.
\label{eq:a3_dyn}
\end{equation}
Here, we note that although only the linear coupling in $h$ is retained here, it generally represents the \emph{leading-order} term of an external drive; higher-order odd contributions may also arise, capturing weak nonlinear bias effects.

The significance of the \(A_3\) normal form lies in its \emph{universality}: a wide range of physical systems exhibiting double-well energy landscapes can be locally approximated by this form through suitable choices of the parameters \(\alpha\) and \(r\). Higher-order corrections may be introduced to account for device-specific nonlinearities; however, provided these corrections remain sufficiently small so as not to alter the non-degenerate nature of the saddle point, they do not modify the essential topology of the potential landscape. In particular, the local dynamical structure in the neighborhood of the barrier maximum—where stochastic transitions originate and Boltzmann sampling is initiated—remains topologically equivalent across implementations. Consequently, the short-time sampling dynamics near the barrier are largely independent of the microscopic device physics, underscoring the $A_3$ normal form as a unifying framework for symmetric bistable systems.

\subsection{Boltzmann Sampling using the $A_3$ Canonical Form}
To evaluate whether Boltzmann-sampling dynamics can be supported by the $A_3$ normal form potential (Eq.~\eqref{eq:A3_Energy}), 
we apply the canonical mapping of Eq.~\eqref{eq:map_eq}, yielding,
\begin{equation}
f(\vartheta) \;=\; \tanh\!\left(\beta h \int^{\vartheta} \frac{d\psi}{g(\psi)} \right),
\label{eq:f-map}
\end{equation}
where, for the $A_3$ normal form,
\begin{equation}
g(\vartheta) = h + \alpha r\,\vartheta - \alpha \vartheta^{3}
\label{eq:g-def}
\end{equation}
At the top of the barrier $\vartheta \sim0$, $g(\vartheta)$ can be approximated as, $g(\vartheta) = h + \alpha r\,\vartheta$. Under these conditions, the effective map can be calculated as,
\begin{equation}
\begin{split}
f(\vartheta)&=\tanh\left(\beta h\int_{0}^{\vartheta} \frac{d\psi}{h+\alpha r\psi}\right)\\\\
&= \tanh\left(\frac{\beta h}{\alpha r} \ln\!\left( 1 + \frac{\alpha r\,\vartheta}{h} \right)\right)\\ \label{eq:A3_map}
\end{split}
\end{equation}
For $\vartheta \rightarrow 0$ ($h \neq 0$), Eq.~\eqref{eq:A3_map} can be approximated as,
\begin{equation}
\begin{split}
s\,=\, f(\vartheta)\,&\approx\, \tanh\left(\beta\vartheta\right)\\\\
 \Rightarrow \quad f'(\vartheta) &=\beta \operatorname{sech}^2(\beta \vartheta) \,=\, \beta (1-s^2) \label{eq:A3_tanh}
\end{split}
\end{equation}

Interestingly, the map in Eq.~\eqref{eq:A3_tanh} exhibits the property
\begin{equation}
\operatorname{sgn}[s] = \operatorname{sgn}[\vartheta],
\end{equation}
where the sign of $\vartheta$ identifies the well (or state) into which the system ultimately settles. This indicates that, for the $A_3$ normal form, the internal state variable itself (or, a its linearly scaled version) approximately samples from a Boltzmann distribution in the vicinity of the barrier region. The fidelity of this approximation depends on the magnitude of higher-order nonlinearities and device-specific departures from the ideal normal form, which influence the effective noise characteristics. Importantly, since a broad class of double-well physical systems can be locally approximated by the $A_3$ normal form near the barrier maximum, their internal state variables may likewise be expected to support approximate Boltzmann sampling, provided the system can be driven sufficiently close to the barrier top. This suggests that classical realizations such as bistable CMOS latches and Duffing-type nonlinear oscillators may be harnessed for approximate Boltzmann sampling under suitable initialization and noise conditions.

Furthermore, we also explore how the $A_3$ normal form connects to the specific example of the oscillator under injection locking for which a well-defined, nearly bias-independent map was obtained in the limiting case of $K_s \rightarrow0$. Starting with
\[
\dot\vartheta = -\gamma\cos\vartheta + K_s\sin(2\vartheta),
\]
we expand about \(\vartheta = 0\) which yields,
\begin{widetext}
\begin{equation}
\begin{aligned}
\dot{\vartheta}
&=
\underbrace{\Big(
2K_s\,\vartheta
-\frac{4}{3}K_s\,\vartheta^{3}
+\frac{4}{15}K_s\,\vartheta^{5}
-\frac{8}{315}K_s\,\vartheta^{7}
+\cdots
\Big)}_{\text{from }K_s\sin(2\vartheta)}
+
\underbrace{\Big(
-\gamma
+\frac{\gamma}{2}\,\vartheta^{2}
-\frac{\gamma}{24}\,\vartheta^{4}
+\frac{\gamma}{720}\,\vartheta^{6}
-\cdots
\Big)}_{\text{from }-\gamma\cos\vartheta} \\[10pt]
&=
\underbrace{\Big(
-\gamma
+ 2K_s\vartheta
-\tfrac{4}{3}K_s\vartheta^{3}
\Big)}_{\text{canonical $A_3$ normal form}}
+
\underbrace{\Big(
\tfrac{4}{15}K_s\vartheta^{5}
-\tfrac{8}{315}K_s\vartheta^{7}
+\cdots
\Big)}_{\text{higher-order coupling terms}}
+
\underbrace{\Big(
\tfrac{\gamma}{2}\vartheta^{2}
-\tfrac{\gamma}{24}\vartheta^{4}
+\tfrac{\gamma}{720}\vartheta^{6}
-\cdots
\Big)}_{\text{nonlinear bias contributions}}.
\end{aligned}
\label{eq:osc_taylor_split_refined}
\end{equation}
\end{widetext}

\noindent
Equation~\eqref{eq:osc_taylor_split_refined} demonstrates that the oscillator dynamics also contain the canonical \(A_3\) normal form supplemented by two families of \emph{higher-order corrections}:  (i) odd-order terms arising from the nonlinear phase response \(\sin(2\vartheta)\); and  
(ii) even-order terms originating from the nonlinear dependence of the external drive \(-\gamma\cos\vartheta\).  Interestingly, these additional contributions render the effective map \(f(\vartheta)\) nearly independent of the bias parameter \(\gamma\) for small barrier heights.

\begin{figure}[h]
    \centering
    \includegraphics[width=1\linewidth]{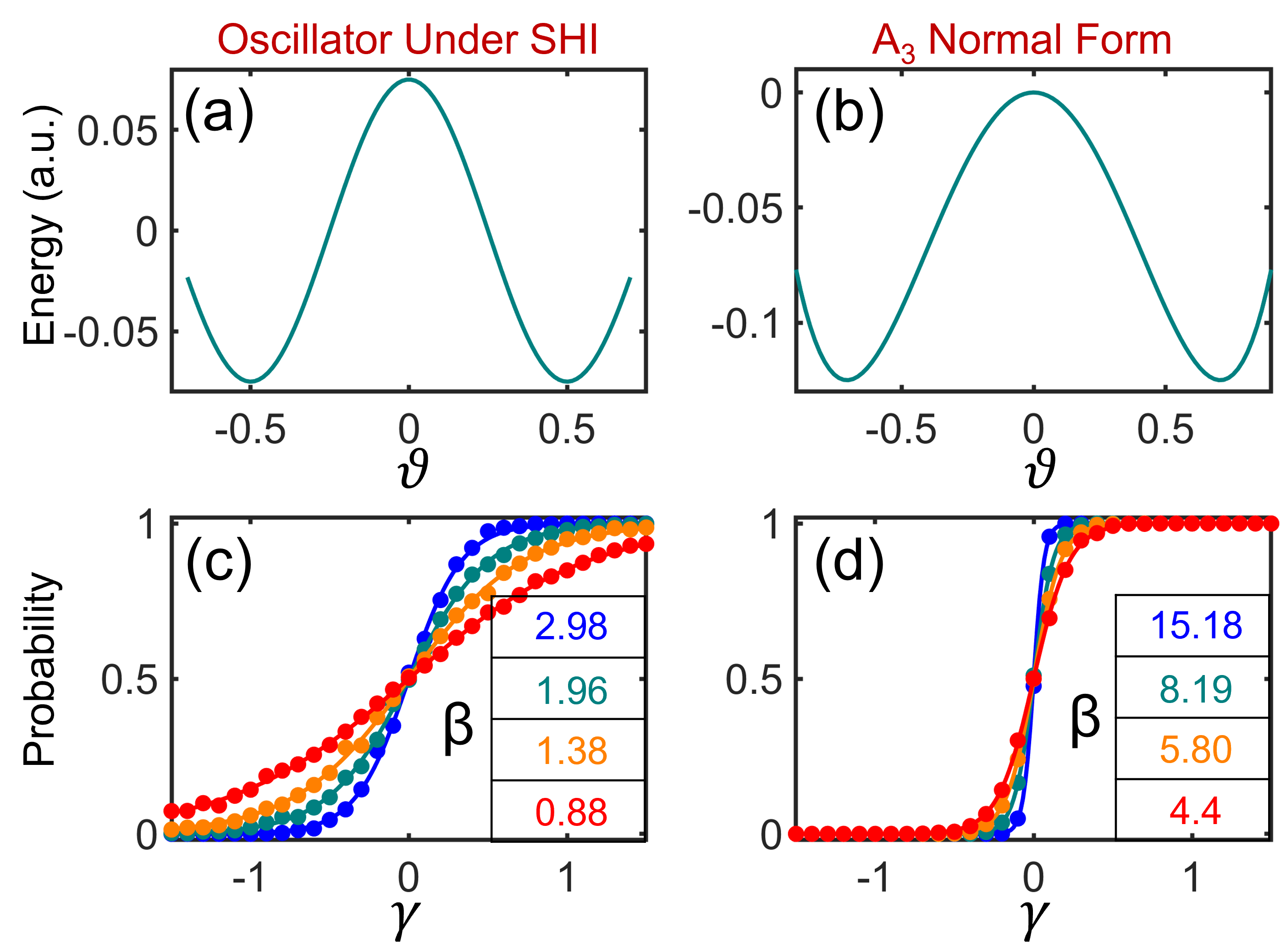}
    \caption{\textbf{Illustrative Double Well Energy Systems:} Energy as a function of the internal state variable for (a) oscillator under SHI (b) $A_3$ normal form. Measured probability distribution ($p$) as a function of synaptic input for: (c) oscillator under SHI; (d) A3 normal form. The simulations have been performed under the following noise conditions:$K_n=0.1$ (blue); $K_n=0.2$ (cyan); $K_n=0.3$ (orange); $K_n=0.4$ (red). Inset shows the extracted inverse temperature $\beta$.}
    \label{fig:DW}
\end{figure}

\section{Simulation Results}

To corroborate the above analysis, we first simulate the stochastic sampling behavior of both the canonical \(A_3\) normal form and an oscillator operating under second-harmonic injection (SHI). 
In both cases, the system dynamics are governed by a gradient-flow equation with additive noise,
\[
d\mu = -\,U'(\mu)\,dt + \sqrt{2K_n}\,dW_t,
\]
where, \(\mu\) represents the internal state variable, \(U(\mu)\) is the potential energy landscape, \(K_n\) denotes the noise intensity, and \(W_t\) is a standard Wiener process. 
Each system is simulated for \(K_n = \{0.1, 0.2, 0.3, 0.4\}\), and the trajectories are numerically integrated using the Euler–Maruyama scheme.

While stochastic sampling benefits from operating with a low energy barrier, as discussed above, reliable retention and readout of the sampled state require a sufficiently large barrier once the "decision" is made. To reconcile these competing requirements, we employ a time-dependent barrier modulation protocol: each stochastic sampling event begins with a shallow barrier that is gradually increased over time. During the low-barrier phase, the system can explore both wells and sample according to the underlying stochastic dynamics. As the barrier is subsequently raised, the minima deepen and the selected state becomes stabilized, enabling robust retention and readout of the sampled outcome. In the $A_3$ normal form (Eq.~\eqref{eq:s_pitch_fork}), this barrier modulation is implemented by increasing the parameter $r$ over time, whereas in the SHI oscillator an analogous effect is achieved by ramping the SHI strength $K_s$.

Figure~\ref{fig:DW} illustrates the double-well potential together with the steady-state probability distributions (symbols) obtained under different noise intensities. 
The solid lines represent fits to the Boltzmann distribution,
\(
p = \frac{1+\tanh(\beta h)}{2}.
\)
In both systems, the simulated probability distributions, computed using 2000 trials, exhibit excellent agreement with the Boltzmann form (\(R^2 > 0.99\)), confirming that they faithfully reproduce the expected stochastic behavior. However, the effective inverse temperature extracted for each device at the same noise intensity ($K_n$) differs, reflecting the influence of each system’s intrinsic curvature and nonlinearity on its effective sampling temperature (as discussed earlier).

\begin{figure}[h]
    \centering
    \includegraphics[width=1\linewidth]{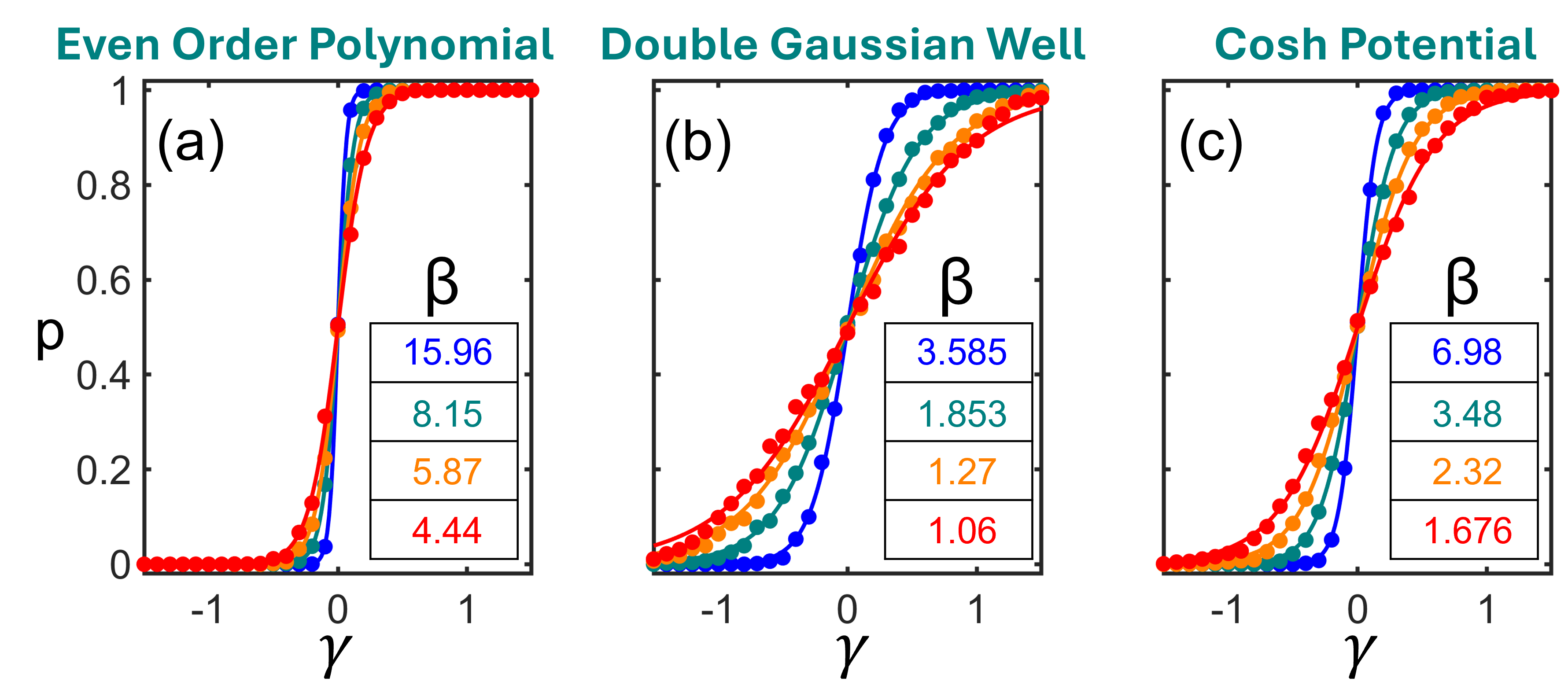}
    \caption{\textbf{Other Double Well Energy Systems:} Measured probability distribution ($p$) as a function of synaptic input for various potential functions: (a) Even order polynomial (b) Double Gaussian Potential (c) Cosh Potential. The exact potentials have been defined in the main text. The simulations have been performed under the following noise conditions:$K_n=0.1$ (blue); $K_n=0.2$ (cyan); $K_n=0.3$ (orange); $K_n=0.4$ (red). Inset shows the extracted inverse temperature $\beta$.}
    \label{fig:General}
\end{figure}

Extending this analysis, Fig.~\ref{fig:General} presents the corresponding results for additional smooth double-well potentials summarized in Table~\ref{tab:potentials}. 
Despite variations in analytic form, all cases yield equilibrium distributions consistent with the Boltzmann law. This finding reinforces the ubiquity of the sampling mechanism.

\section{Conclusion}

This work establishes a ubiquitous mechanism underlying stochastic sampling in bistable, double-well energy systems when initialized near the barrier top. By adopting a topological perspective grounded in Morse theory, we leverage the fact that, in the vicinity of a nondegenerate saddle point, any smooth, even double-well potential can be locally reduced to a quartic ($A_3$) normal form. Within this locally universal regime, the combined action of gradient flow, thermal noise, and synaptic bias produces a characteristic short-time relaxation that exhibits a $\tanh(\cdot)$-like response. This response preserves the functional structure of Boltzmann sampling, even though the effective temperature scaling may depend on device-specific parameters such as curvature and noise strength. As a result, the emergent Boltzmann-like behavior near the barrier top is not a device-dependent artifact, but a generic consequence of the local topology of bistable energy landscapes.

These insights motivate and justify an alternative sampling paradigm—\emph{synchronous sampling}—which contrasts with traditional asynchronous p-bit operation. In asynchronous schemes, Boltzmann statistics arise from the distribution of dwell times during continuous stochastic switching between wells, which depend on the ratio of the energy barrier height to the thermal noise intensity. In synchronous sampling, by contrast, each p-bit is periodically driven to the barrier top and allowed to relax for a short, controlled time under the influence of bias and noise. This clocked operation yields a discrete sampling event whose statistics are governed by the same universal short-time dynamics, while enabling deterministic timing, parallelism, and architectural synchronization across large arrays.

From an applications perspective, this framework provides a unifying theoretical foundation for evaluating and comparing a wide range of physical implementations—including CMOS bistable latches, oscillators under harmonic injection, and magnetic tunnel junctions—as probabilistic computing primitives. By decoupling the universality of the sampling mechanism from the microscopic details of device physics, our results bridge device-level dynamics and algorithmic-level functionality, and offer a principled pathway toward scalable, noise-driven probabilistic computing architectures based on synchronous operation.

\section*{Acknowledgments}
\noindent This work was supported by National Science Foundation grant \#2433871. \\

\textbf{DATA AVAILABILITY:} The data that support the findings of this study are available from the corresponding author upon reasonable request.

\appendix

\section{\\Formulating the Map $s=f(\vartheta)$}
\label{ap:s_map}
For gradient flow dynamics given by $\dot\vartheta=g(\vartheta)$, the corresponding mapping to $s$--where $s=\tanh(\beta h t)$ can be determined as follows:

\begin{equation}
    \begin{split}
        &\,\,\dot{s}=f'(\vartheta).\dot{\vartheta}\\\\
    \end{split}
\end{equation}
Using the relationship: $\dot s=\beta h (1-s^2)$,
\begin{equation}
    \begin{split}
        \quad & \beta h\left(1- s^2\right)=f'(\vartheta).g(\vartheta)\\\\
        \Rightarrow \quad & \beta h\left(1- f(\vartheta)^2\right)=f'(\vartheta).g(\vartheta)\\\\
        \Rightarrow \quad &\beta h\int \frac{d\vartheta}{g(\vartheta)}\,=\, \int \frac{df(v)}{1- f(\vartheta)^2}\\\\ \label{eq:s_map}
        \Rightarrow \quad & f(\vartheta)\,=\,\tanh\left(\beta h \int \frac{d\vartheta}{g(\vartheta)}\right)\,+\,C
    \end{split}
\end{equation}

$C=0$ if $s=0$ when $\vartheta=0$ which is condition for a valid mapping. 

\section{\\Mean First-Passage Time for the Logistic Drift Model}
\label{ap:delta_t}

We assume $\Delta t$ to be the Mean First-Passage Time (MFPT) and derive a closed-form integral expression for the MFPT to reach
the symmetric absorbing thresholds $\pm z_{\mathrm{th}}$ for the 1D stochastic process
\begin{equation}
dz \;=\; b(z)\,d\tau \;+\; \sqrt{2K_n}\,dW_\tau,
\qquad
b(z)=\beta\,h\,(1-z^2),
\label{eq:SDE_appendix}
\end{equation}

Let $\tau$ denote the first exit time from the interval $(-z_{\mathrm{th}},\,z_{\mathrm{th}})$:
\begin{equation}
\tau \;=\; \inf\{\, t>0 \;:\; z(t)\notin (-z_{\mathrm{th}},z_{\mathrm{th}})\,\}.
\end{equation}
The MFPT starting at $z$ is
\begin{equation}
u(z)\;:=\;\mathbb{E}_{z}[\tau].
\end{equation}
By standard backward Kolmogorov/FP theory, $u(z)$ solves the linear ODE
\begin{equation}
b(z)\,u'(z)\;+\;K_n\,u''(z)\;=\;-1,
\qquad z\in(-z_{\mathrm{th}},z_{\mathrm{th}}),
\label{eq:backward_ode}
\end{equation}
with absorbing boundary conditions
\begin{equation}
u(-z_{\mathrm{th}})=u(z_{\mathrm{th}})=0.
\label{eq:BCs}
\end{equation}

\noindent For $z_{\mathrm{th}}\ll 1$ the drift is nearly constant $b(z)\approx k h$. 

The MFPT then reduces to the classic drifted-Brownian result
\begin{equation}
\Delta t \sim \mathbb{E}[\tau]\;\approx\;\frac{z_{\mathrm{th}}}{k h}\,\tanh\!\Big(\frac{k h\,z_{\mathrm{th}}}{2K_n}\Big)
\end{equation}

If $\tfrac{1}{2}k h\,z_{\mathrm{th}}\ll K_n$, the MFPT behaves diffusively:
\begin{equation}
\Delta t \sim \mathbb{E}[\tau]\;\approx\;\frac{z_{\mathrm{th}}^{2}}{2K_n}
\end{equation}

It can be observed that in this regime (when $h\ll1$ is very small), $\Delta t$ is independent of the synaptic input $h$.

If $\tfrac{1}{2}k h\,z_{\mathrm{th}}\gg K_n$, then
\begin{equation}
\Delta t \sim \mathbb{E}[\tau]\;\approx\;\frac{z_{\mathrm{th}}}{k h}
\end{equation}

\section{Even polynomials that yield a Double Well Energy Landscape}
\label{ap:DW}
\noindent Consider the even polynomial potential:
\begin{equation}
U(x)
=\sum_{k=2}^{n} a_{2k}x^{2k}-b\,x^{2},
\qquad a_{2k}>0,\; b>0.
\end{equation}
The first derivative is,
\begin{equation}
U'(x)
=\sum_{k=2}^{n} (2k)a_{2k}x^{2k-1}-2b\,x
=x\,Q(x^{2}),
\end{equation}
where,
\begin{equation}
Q(y)
=\sum_{k=2}^{n} (2k)a_{2k}y^{k-1}-2b,
\qquad y=x^{2}\ge 0.
\end{equation}

\noindent Since $U(x)$ is even (only even powers appear), we have $U(-x)=U(x)$ and hence $U'$ is odd: $U'(-x)=-U'(x)$. 
Therefore, critical points occur in symmetric pairs about the origin, and it suffices to analyze $x\ge 0$; the behavior on $(-\infty,0]$ follows by symmetry. 
Moreover, writing $U'(x)=x\,Q(x^{2})$ shows that the auxiliary variable $y=x^{2}$ is nonnegative, so $Q$ needs only be considered on $[0,\infty)$. \\

\noindent We note that,
\[
Q(0)=-2b<0, 
\qquad 
\lim_{y\to\infty}Q(y)=+\infty.
\]
and
\[
Q'(y)
=\sum_{k=3}^{n}(2k)(k-1)a_{2k}y^{k-2}>0
\quad \text{for all } y>0,
\]
since $a_{2k}>0$. Therefore, $Q(y)$ is \emph{strictly increasing} on $[0,\infty)$ and hence possesses exactly one positive root $y^{*}>0$.\\

\noindent From $U'(x)=xQ(x^{2})$, the stationary points are:
\[
x=0, \qquad x=\pm \sqrt{y^{*}}.
\]
Thus, $U$ has three critical points in total.\\

\noindent Differentiating again,
\begin{equation}
U''(x)=Q(x^{2})+2x^{2}Q'(x^{2}).
\end{equation}
At $x=0$,
\[
U''(0)=Q(0)=-2b<0,
\]
so $x=0$ is a strict \emph{local maximum} (the barrier top).  
At $x=\pm \sqrt{y^{*}}$, using $Q(y^{*})=0$,
\[
U''(\pm \sqrt{y^{*}})=2y^{*}Q'(y^{*})>0,
\]
so both are strict \emph{local minima}.

Furthermore, as $|x|\to\infty$, $U(x)\sim~a_{2n}x^{2n} \rightarrow \infty$ implying that $\pm y^{*}$ are global minimia. 

Consequently, even polynomials of the above form represent a classic double well energy landscape.

\clearpage
\def\bibsection{\section*{References}}  
\bibliography{Arxiv}

\end{document}